\documentclass[superscriptaddress,floatfix,showpacs,preprint] {revtex4}
\usepackage{amsmath,amssymb,epsfig,psfig,color,pstricks,bm,graphics,alltt}

\newcommand{\BK}{{\ensuremath{\bm{k}}}}
\newcommand{\BG}{{\ensuremath{\bm{G}}}}
\newcommand{\BS}{{\ensuremath{\bm{\Sigma}}}}

\begin{document}
  \title{Momentum-resolved spectral functions of SrVO$_3$
    calculated~by~LDA+DMFT}

  \author{I.~A.~Nekrasov}
  \affiliation{Institute for Electrophysics, Russian Academy of
    Sciences, Ekaterinburg, 620016, Russia}
  \affiliation{Theoretical Physics III, Center for Electronic
    Correlations and Magnetism, University of Augsburg, 86135
    Augsburg, Germany}

  \author{K.~Held}
  \affiliation{Max-Planck Institute for Solid State Research,
    Heisenbergstr.~1, 70569 Stuttgart, Germany}

  \author{G.~Keller}
  \affiliation{Theoretical Physics III, Center for Electronic
    Correlations and Magnetism, University of Augsburg, 86135
    Augsburg, Germany}

  \author{D.~E.~Kondakov}
  \affiliation{Institute of Metal Physics, Russian Academy of
    Sciences, Ekaterinburg, 620219, Russia}

  \author{Th.~Pruschke}
  \affiliation{Institute for Theoretical Physics, University of
    G\"ottingen, Tammannstr.\ 1, 37077 G\"ottingen, Germany}

  \author{M.~Kollar}
  \affiliation{Theoretical Physics III, Center for Electronic
    Correlations and Magnetism, University of Augsburg, 86135
    Augsburg, Germany}

  \author{O.~K.~Andersen}
  \affiliation{Max-Planck Institute for Solid State Research,
    Heisenbergstr.~1, 70569 Stuttgart, Germany}

  \author{V.~I.~Anisimov}
  \affiliation{Institute of Metal Physics, Russian Academy of
    Sciences, Ekaterinburg, 620219, Russia}

  \author{D.~Vollhardt}
  \affiliation{Theoretical Physics III, Center for Electronic
    Correlations and Magnetism, University of Augsburg, 86135
    Augsburg, Germany}

  \begin{abstract}
    
    {{LDA+DMFT, the merger of density functional theory in the
        local density approximation and dynamical mean-field theory,
        has been mostly employed to calculate ${\BK}$-integrated
        spectra accessible by photoemission spectroscopy.  In this
        paper, we calculate ${\BK}$-{\it{resolved}} spectral functions
        by LDA+DMFT. To this end, we employ the $N$th order muffin-tin
        (NMTO) downfolding to set up an effective low-energy
        Hamiltonian with three $t_{2g}$ orbitals.  This downfolded
        Hamiltonian is solved by DMFT yielding ${\BK}$-dependent
        spectra. Our results show renormalized quasiparticle bands
        over a broad energy range from -0.7~eV to +0.9~eV with small
        ``kinks'', discernible in the dispersion below the Fermi energy.}}
  \end{abstract}

  \maketitle

  \section{Introduction}\label{introduction}
  
  Transition metal oxides show a diversity of challenging physical
  phenomena, including superconductivity, metal-insulator transitions,
  and colossal magnetosresistance, and are therefore at the center of
  modern solid state research. Electrons in many of these materials
  are strongly correlated due to a large ratio of Coulomb interaction
  to bandwidth $U/W$, resulting in complicated many-electron physics
  which makes realistic calculations rather difficult. In particular,
  conventional bandstructure calculations, e.g., in the local density
  approximation (LDA)\cite{JonesGunn}, fail because these effective
  one-particle approaches do not contain many-body physics like the
  formation of Hubbard bands, quasiparticle renormalization, and
  lifetime effects. In this respect LDA+DMFT, the recent
  merger~\cite{AnLi,Held01,heldpsik,likako,PT} of LDA with the many-body dynamical
  mean-field theory
  (DMFT)~\cite{MetzVoll89,vollha93,pruschke,georges96}, is a promising
  new approach which includes many-body aspects into realistic
  calculations. It has been successfully applied, in particular to
  calculate the total (${\BK}$-integrated) spectra of
  transition-metal oxides like LaTiO$_3$~\cite{AnLi,LDADMFTTMO2},
  V$_2$O$_3$~\cite{Held01a,Laad02b,Wannier},
  Sr(Ca)VO$_3$~\cite{LiebschSrVO3,Nekrasov03,Pavarini03,Sekiyama03,Wannier,Nekrasov05x,Pavarini05},
  LiV$_2$O$_4$~\cite{Nekrasov02},
  Ca$_{2-x}$Sr$_x$RuO$_4$~\cite{LDADMFTTMO1, Anisimov01},
  CrO$_2$~\cite{Laad02a}, but also of Ni~\cite{Lichtenstein01},
  Fe~\cite{Lichtenstein01}, and $f$-electron systems like
  Pu~\cite{SAVRASOV,SAVRASOV2} and
  Ce~\cite{Zoelfl01,McMahan01,McMahan03,Haule05,McMahan05,Georges05}.
  
  LDA+DMFT calculations for transition metal oxides have mostly been
  restricted to the $d$-bands around the Fermi energy, employing a
  simplified calculational scheme based on the LDA density of states
  (DOS) which holds for cubic systems~\cite{Held01}.  Calculations
  with the full LDA Hamiltonian, including all {\em spdf} valence
  orbitals in the DMFT have been performed for
  Pu~\cite{SAVRASOV,SAVRASOV2} and
  Ce~\cite{Zoelfl01,McMahan01,McMahan03,Haule05}.  Since the
  O$_{2p}$-Me$_{3d}$ overlap is considerable for transition metal
  oxides, a full LDA Hamiltonian calculation should also take into
  account the rather large~\cite{McMahan,Hybertsen,Knotek}
  oxygen-vanadium and oxygen-oxygen Coulomb interactions $U_{pd}$ and
  $U_p$.
  
  While a large number of interacting orbitals makes LDA+DMFT
  calculations with the full $spd$ Hamiltonian difficult, they are
  feasible for the effective $d$-bands around the Fermi energy.  To
  this end, a clear definition of the effective Hamiltonian for
  energies near the Fermi energy is mandatory. An accurate
  construction of this effective Hamiltonian is possible by the
  downfolding procedure for third generation muffin-tin orbitals
  (NMTOs) \cite{MTO3} and has recently been employed in the LDA+DMFT
  context by Pavarini {\em et al.} \cite{Pavarini03,Pavarini05}.
  Furthermore, Anisimov~{\em et al.}~\cite{Wannier} recently proposed
  a projection scheme of the Bloch functions onto a Wannier functions
  basis to obtain a few-orbital Hamiltonian.  Such a downfolded or
  projected Hamiltonian is also required to calculate
  ${\BK}$-resolved spectra.
  
  Due to its simple crystal structure (cubic perovskite) and the
  3$d^{1}$ electronic configuration the transition metal oxide
  SrVO$_3$ is ideal for testing new theoretical methods for the
  realistic modeling of correlated materials.  SrVO$_{3}$ is a
  strongly correlated metal with a pronounced lower Hubbard and
  quasiparticle peak in the photoemission spectra
  (PES)~\cite{Fujimori92a,Maiti01,Sekiyama02} as well as a pronounced
  quasiparticle and upper Hubbard band in the x-ray absorption
  spectrum~\cite{Inoue94}. After the substitution of Sr by Ca,
  PES~\cite{Fujimori92a} and Bremsstrahlung isochromat spectra
  (BIS)~\cite{Morikawa95} originally suggested the onset of a
  Mott-Hubbard metal-insulator transition. By contrast, thermodynamic
  properties (Sommerfeld coefficient, resistivity, and paramagnetic
  susceptibility)~\cite{old_experiments} did not show significant
  effects upon Ca doping. An attempt to describe electronic properties
  of SrVO$_3$ by DMFT was made by Rozenberg {\em et al.}
  \cite{Rozenberg96,Makino98} for the one-band Hubbard model using
  phenomenological parameters.  Recently, the puzzling discrepancy
  between spectroscopic and thermal properties has been solved by new
  bulk-sensitive PES~\cite{Maiti01,Sekiyama02,Eguchi05}, showing
  similar spectra for CaVO$_{3}$ and SrVO$_{3}$, in agreement with the
  thermodynamic results.  This was confirmed theoretically by LDA+DMFT
  calculations~\cite{Nekrasov03,Pavarini03,Sekiyama03,Wannier,Nekrasov05x,Pavarini05}.
  {{In this paper we present LDA+DMFT(QMC) calculations for
      SrVO$_{3}$ based on a NMTO downfolded effective Hamiltonian for
      three orbitals of $t_{2g}$ symmetry crossing the Fermi energy.
      {}From this we calculate ${\BK}$-resolved spectral functions and
      ARPES spectra. Recently angular-resolved photoemission
      spectroscopy (ARPES) on SrVO$_3$ was performed
      \cite{Yoshida05}; the data from Fujimori's group
      \cite{Yoshida05} allow for the direct observation of the
      quasiparticle mass renormalization and band edge.}}
  
  The paper is organized as follows.  In Section~\ref{LDA} we briefly
  discuss the crystal structure and calculate the effective $t_{2g}$
  Hamiltonian for SrVO$_3$ by LDA/NMTO. In Section~\ref{DMFT}, we
  discuss and compare two LDA+DMFT(QMC) calculations for SrVO$_3$,
  based on this effective $t_{2g}$ Hamiltonian and a simplified
  treatment using the DOS only.  Finally, in Section~\ref{ARPES} the
  LDA+DMFT(QMC) calculated {{self-energy on the real axis
      $\Sigma(\omega)$}} and ${\BK}$-resolved spectral functions for
  SrVO$_{3}$ are presented. The paper is summarized in Section
  \ref{conclusion}.

  \section{Construction of few-orbital Hamiltonians}
  \label{LDA}
  
  Starting point of a first principle calculation is usually the
  crystal structure. In our case, SrVO$_3$ is a perovskite with an
  ideal cubic Pm$\bar 3$m~\cite{Rey90} symmetry, containing one V ion
  in the unit cell. This implies that the main structural element, the
  VO$_6$ octahedron, is not distorted. The electronic configuration is
  3$d^1$, which follows from the formal oxidation V$^{4+}$.  Due to
  the cubic symmetry, the $d$ orbitals split into two sets: three
  $t_{2g}$ and two $e_g$ orbitals.  In our case of an octahedral
  coordination with oxygen, the three-fold degenerated $t_{2g}$
  orbitals are lower in energy than the two-fold $e_g$ orbitals.
  Since these $t_{2g}$ and $e_g$ bands do not overlap we will later
  restrict our calculation to an effective Hamiltonian with three
  $t_{2g}$ orbitals filled with one electron per site.
  
\begin{figure}
    \begin{center}
      \includegraphics[clip=true,width=0.5\textwidth,angle=270]{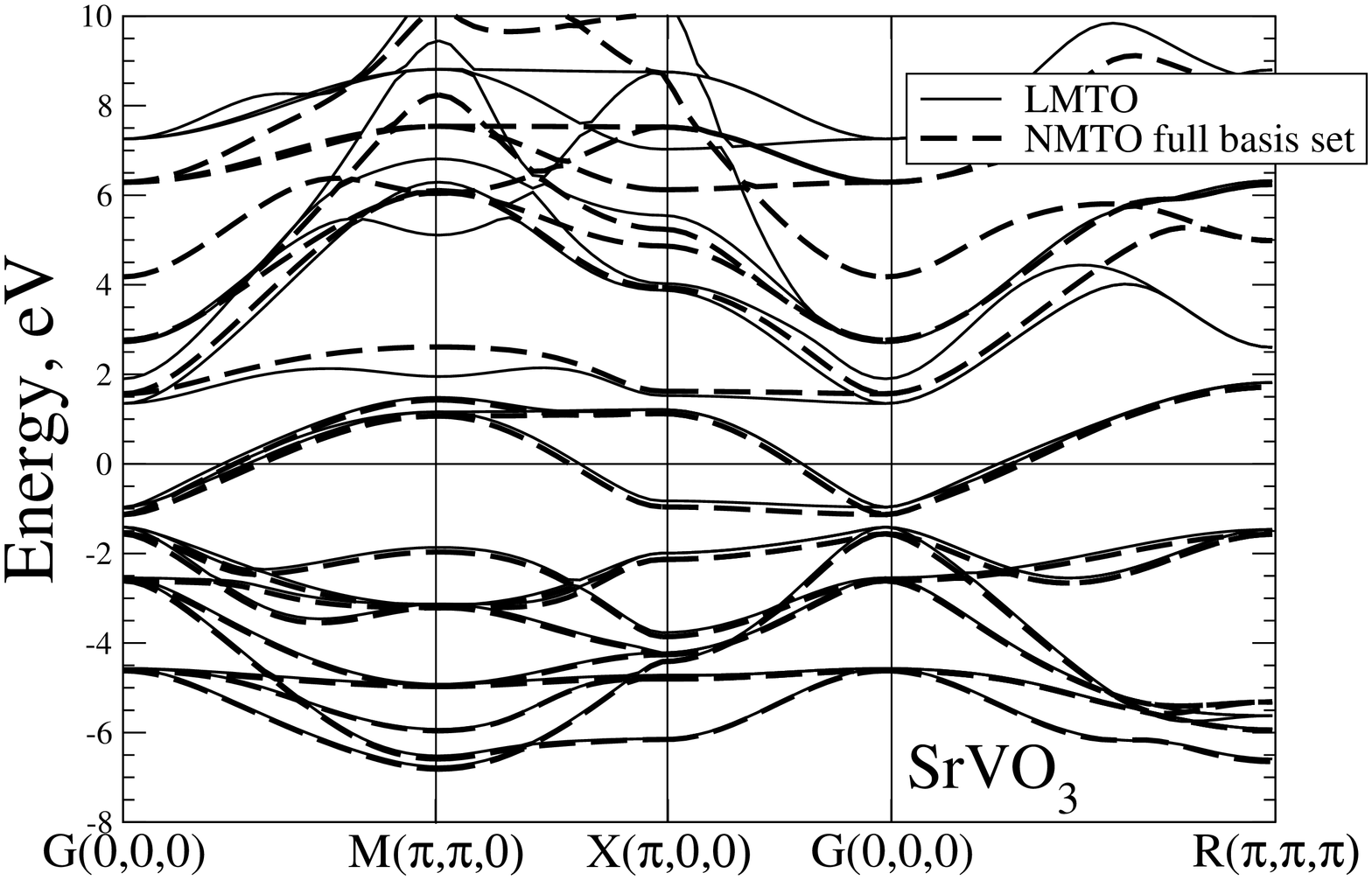}
    \end{center}
    \caption{Comparison of the LDA band structure  of SrVO$_3$ calculated by
      LMTO (thin solid line) and  NMTO  (dashed line)
      for the full orbital basis set. 
      Here, and in the following figures,
      the Fermi energy corresponds to zero energy.}
    \label{lmto_nmto_bands}
  \end{figure}

  For the LDA band structure calculations of SrVO$_3$ we first
  employed the LDA-LMTO(ASA) code version 47 which uses the basis of
  nonorthogonal linearized muffin-tin orbitals (LMTO; 2nd generation)
  in the atomic sphere approximation (ASA)~\cite{LMTO}.  Thereby, the
  partial waves were expanded to linear order in energy around the
  center of gravity of the filled part of the bands. The results are
  presented by thin solid lines in Fig.\ \ref{lmto_nmto_bands} and
  show 2$p$ oxygen bands below -1.5~eV, three $t_{2g}$ bands at the
  Fermi energy between -1.5~eV and 1.5~eV, and $e_g$ bands between
  1.5~eV and 6~eV.  The other bands of our orbital basis set
  [O($3s,3d$), V($4s,4p$), Sr($5s,5p,4d,4f$)] are empty and lie far
  above the Fermi level~\cite{comment}.
  
  Secondly, with the same basis set we employed the third generation
  MTO, also known as $N$-th order muffin tin orbitals (NMTO)
  \cite{MTO3}. We expanded the MTO orbitals around the three points:
  -2.72~eV, 0.68~eV, and 6.8~eV. Here and in the following, all
  energies are measured relative to the Fermi energy at $0\,$eV. The
  NMTO results are shown as dashed lines in Fig.\ 
  \ref{lmto_nmto_bands} and almost coincide with LMTOs in the region
  of interest, i.e., O2$p$ and V3$d$.  The NMTO bands are found to be
  slightly lower in energy which is not surprising since 3rd
  generation MTOs have the proper energy dependence in the
  interstitial region and, moreover, more expansion points ($N+1=4$)
  for the wave function than LMTOs where $N=1$ (linear approach). For
  the high-lying empty bands, LMTO and NMTO bands are quite different;
  the NMTO bands are again lower in energy. As the third NMTO
  expansion point (6.8~eV) is in this region, we expect NMTOs to be
  more precise in this region than LMTOs, which are linearized at
  energies corresponding to the center of gravity of the {\em filled}
  parts of the bands. Hence, the LMTO expansion points are below the
  Fermi energy, far away from these high-lying empty bands. Moreover,
  the 2nd generation LMTOs have vanishing kinetic energy in the
  interstitial region.
  
\begin{figure}
    \begin{center}
      \includegraphics[clip=true,width=0.5\textwidth,angle=270]{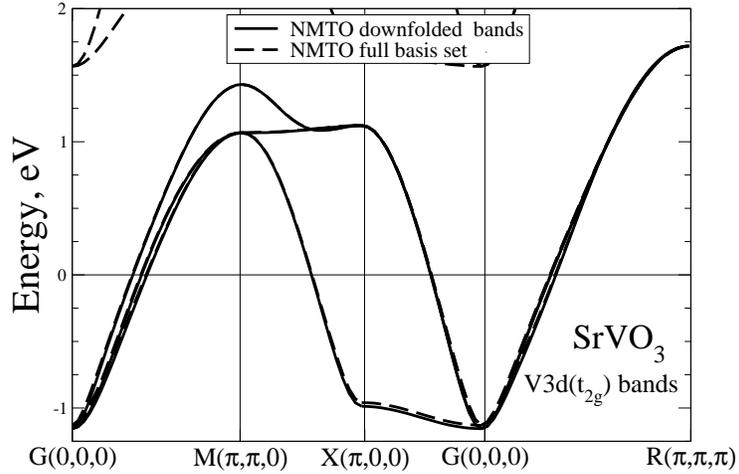}
    \end{center}
    \caption{Comparison of the NMTO  downfolded  $t_{2g}$  bands (full line)
      with  NMTO for the full orbital basis
      set (dashed line).}
    \label{nmto_dnf_bands}
  \end{figure}

  A particular advantage of NMTOs is the possibility of calculating an
  effective (downfolded) Hamiltonian $\hat H^{\text{eff}}({\BK})$,
  confined to a reduced set of orbitals in a reduced window of
  energies. In the case of SrVO$_3$, the $t_{2g}$ subset of the V 3$d$
  orbitals is of particular interest as discussed above.  Hence, we
  downfolded \cite{MTO3} to a $3 \times 3$ NMTO Hamiltonian $\hat
  H^{\text{eff}}({\BK})$ describing the three $t_{2g}$ orbitals. For
  optimizing the energy window w.r.t. these orbitals, we chose two MTO
  expansion points, $\epsilon_0=0.41\,$eV and $\epsilon_1=0.95\,$eV,
  at the energy region of the $t_{2g}$ bands.  At these energies, the
  NMTOs span exactly the LDA eigenfunctions.  Fig.\ 
  \ref{nmto_dnf_bands} shows the eigenvalues of
  $\hat{H}^{\text{eff}}({\BK})$ along some high-symmetric directions
  in comparison with the
  NMTO results using the full orbital basis of Fig.\ 
  \ref{lmto_nmto_bands}. {}From the good agreement we conclude that
  $\hat H^{\text{eff}}({\BK})$ describes the three $t_{2g}$ bands well.
  The slight discrepancy at the bottom of the band could have been
  avoided by choosing a smaller value of $\epsilon_0$.  If we increase
  the number of these mesh points $\epsilon_i$, the Hilbert space
  spanned by these NMTOs will converge to that spanned by the $t_{2g}$
  Wannier functions; the orthogonalization of these NMTOs will yield
  localized Wannier functions.
  
\begin{figure}
    \begin{center}
      \includegraphics[clip=true,width=0.5\textwidth,angle=270]{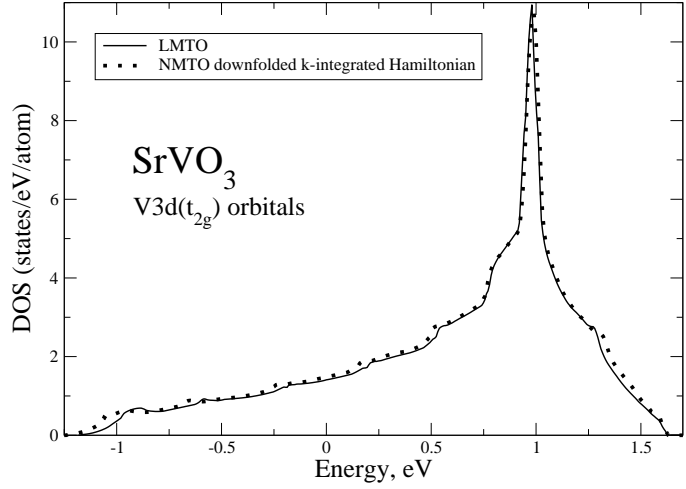}
    \end{center}
    \caption{Comparison of the $t_{2g}$ DOS  calculated
      (i) by  LMTO as explained in the text (thin solid line) and
      (ii) by integrating the down-folded
      NMTO  Hamiltonian over the Brillouin zone (doted line).}
    \label{lmto_nmto_dos}
  \end{figure}

  Fig.\ \ref{lmto_nmto_dos} compares the DOS of $\hat
  H^{\text{eff}}({\BK})$ obtained via tetrahedron integration
  \cite{Lambin84} with the LMTO DOS. A minor difference to earlier
  calculations \cite{Nekrasov03,Nekrasov05x} is that we used an
  orthogonal representation of the LMTO method in Refs.\ 
  \cite{Nekrasov03,Nekrasov05x}, neglecting the so-called combined
  correction term.  Because of this Refs.\ 
  \cite{Nekrasov03,Nekrasov05x} yield a slightly different $t_{2g}$
  bandshape with a discernibly reduction of the sharp peak at $\approx
  1\,$eV.  These differences are, however, small and unimportant for
  the final LDA+DMFT results.  For the LMTO DOS of Fig.\ 
  \ref{lmto_nmto_dos}, we downfolded the band structure onto the
  $t_{2g}$ states which, due to the oxygen 2$p$-$t_{2g}$
  hybridization, also have a contribution between -7~eV and -2~eV.
  Vice versa, downfolding to O-2$p$ states gives a contribution around
  the Fermi energy.  To obtain the {\em effective} $t_{2g}$ orbitals
  at the Fermi energy (which have primarily $t_{2g}$ character with a
  small $2p$ admixture) we truncated the $t_{2g}$ contribution in the
  oxygen region between -7~eV and -2~eV and renormalized the orbitals
  so that one has again one electron per site and orbital.
  Fig.~\ref{lmto_nmto_dos} shows that the DOS calculated by this
  procedure resembles the downfolded NMTO DOS well.  In particular,
  both DOSes have the same features and bandwidth.  The agreement with
  the NMTO DOS of \cite{Pavarini03} is also very good.

  \section{LDA+DMFT calculations using downfolding and Hilbert
    transform}
  \label{DMFT}
  
  In this Section, we will use two different methods to construct the
  non-interacting, i.e., kinetic energy part, of the three-band
  many-body problem: the NMTO downfolded $t_{2g}$ Hamiltonian $\hat
  H^{\text{eff}}({\BK})$ and the LMTO DOS of Fig.~\ref{lmto_nmto_dos}.
  This part of the Hamiltonian is then complemented by a local Coulomb
  interaction:
  \begin{eqnarray}
    \hat{H} =\hat{H}_{0}^{\text{eff}}&+&U\sum_{m}\sum_{i}
    \hat{n}_{im\uparrow}\hat{n}_{im\downarrow }\label{H} 
    \nonumber \\ &+&
    \;\sum_i\sum_{m\neq m'}\sum_{\sigma \sigma'}\;
    (U'-\delta_{\sigma \sigma'}J)\;
    \hat{n}_{im\sigma}\hat{n}_{im'\sigma'}.
    \label{Hamiltonian}
  \end{eqnarray}
  Here, the index $i$ enumerates correlated lattice sites, $m$ denotes
  orbitals, and $\sigma$ the spin. $\hat H_{0}^{\text{eff}}$ is a
  one-particle Hamiltonian generated from the LDA band structure where
  an averaged Coulomb interaction is subtracted to avoid double
  counting of the Coulomb interaction~\cite{AnLi,Held01}.  The local
  intra-orbital Coulomb repulsion is denoted by $U$ and the Hund's
  exchange coupling by $J$. Rotational invariance then fixes the local
  inter-orbital Coulomb repulsion $U^\prime=U-2J$, see, e.g., \cite{Zoelfl00}. For
  three orbitals, $U^\prime$ equals the averaged Coulomb interaction
  $\bar U$~\cite{LDADMFTTMO2,Held01}.
  
  The Hamiltonian~(\ref{Hamiltonian}) is then solved by the recently
  developed LDA+DMFT approach~\cite{AnLi} (for introductions see
  Refs.~\cite{Held01,PT}, for reviews see
  Refs.~\cite{heldpsik,likako}).  In this approach the solution
  of~(\ref{Hamiltonian}) is obtained by the dynamical mean-field
  theory (DMFT)~\cite{vollha93,pruschke,georges96}, a non-perturbative
  many-body method based on the $d=\infty$ limit~\cite{MetzVoll89}.
  
  In this paper, $\hat H_{0}^{\text{eff}}$ will be the NMTO downfolded
  (and symmetrically orthonormalized) Hamiltonian of Section
  \ref{LDA}. The double counting correction is not relevant here since
  we consider only the three correlated $t_{2g}$
  orbitals~\cite{LDADMFTTMO2,Held01}.  To calculate Coulomb
  interaction parameters appearing in~(\ref{Hamiltonian}) we
  previously~\cite{Nekrasov03,Nekrasov05x} employed the constrained LDA
  method~\cite{Gunnarsson89}, yielding an orbitally averaged Coulomb
  repulsion $\bar U$=3.55~eV and a Hund's exchange coupling
  $J$=1.0~eV.

  In our LDA+DMFT calculations, we self-consistently solve the
  auxiliary DMFT impurity problem~\cite{vollha93,pruschke,georges96}
  by multi-band quantum Monte Carlo (QMC) simulations~\cite{QMC}
  together with the ${\BK}$-integrated Dyson equation:
  \begin{eqnarray}
    \bm{G}(\omega)=\int\limits_{\text{BZ}}\!d{\BK}\;[\omega+\mu -
    \bm{\Sigma}(\omega) - \bm{h}^{\text{eff}}_{0}({\BK})]^{-1}.
    \label{Ham_intg}
  \end{eqnarray}
  Here, $\BG(\omega)$, $\BS(\omega)$, 
  and $\bm{h}^{\text{eff}}_{0}({\BK})$ are $3 \times 3$ matrices in orbital space,
  denoting the Green function, self-energy, and the downfolded NMTO
  Hamiltonian $\hat{H}^{\text{eff}}_{0}$ in reciprocal space,
  respectively; $\mu$ is the chemical potential. Since QMC is
  formulated on the imaginary axis, we employed Eq.\ (\ref{Ham_intg})
  for Matsubara frequencies and analytically continued $G(\omega)$ to
  real frequencies by means of the maximum entropy method \cite{MEM}.
  
  In our previous calculations \cite{Nekrasov03,Nekrasov05x}, we used a simplified
  scheme based on the LDA DOS only. Within a cubic symmetry, the local
  DMFT self-energy becomes diagonal and even orbital-independent:
  $\Sigma_{mm'\sigma\bar\sigma}(\omega)=\delta_{m m'}
  \delta_{\sigma\sigma'} \Sigma(\omega)$.  Then, the Green functions
  $G_(\omega)$ of the lattice problem can be expressed via the Hilbert
  transform of the LDA DOS $N^0(\epsilon)$:
  \begin{equation}
    G(\omega)=\int d\epsilon \frac{N^{0}(\epsilon
    )}{\omega+\mu-\epsilon-\Sigma(\omega)+i\eta},
    \label{intg}
  \end{equation}
  instead of Eq.\ (\ref{Ham_intg}).
  
  In Fig.~\ref{qmc_nmto_dos}, we present a comparison between
  one-particle LDA+DMFT(QMC) spectra for SrVO$_3$ obtained by using
  Eq.\ (\ref{intg}) with the Vanadium $t_{2g}$ LDA DOS (thin solid
  line in Fig.~\ref{lmto_nmto_dos}; calculated as described in Section
  \ref{LDA}) and Eq. (\ref{Ham_intg}) with
  $\bm{h}^{\text{eff}}_{0}({\BK})$.  Both methods give the same
  results, as is to be expected for a cubic system.  One can see the
  generic ``three-peak'' spectrum of a strongly correlated metal: the
  quasiparticle peak slightly above the Fermi energy, and lower and
  upper Hubbard bands to the left and right.  The results presented
  here agree well with those reported in
  Refs.~\cite{Nekrasov03,Pavarini03,Sekiyama03,Wannier,Nekrasov05x,Pavarini05}. {{The LDA+DMFT calculations of
      Ref.\ \cite{LiebschSrVO3}, with a focus on bulk surface differences,
      used a somewhat lower Coulomb interaction
      $\bar{U}=U-2J=2.6,2.9$~eV.}}

  \section{Calculation of ${\BK}$-resolved spectra}
  \label{ARPES}
  
  The purpose of this paper is to calculate the ${\BK}$-resolved
  spectral function $A({\BK},\omega)$ for SrVO$_3$ within the
  LDA+DMFT(QMC) scheme. Here,
  \begin{eqnarray}
    A({\BK},\omega)&=&-\frac{1}{\pi}{\rm{Im Tr}} \BG({\BK},\omega)
    \label{aofomega}
  \end{eqnarray}
  is determined by the diagonal elements of the Green function matrix in orbital space
  \begin{eqnarray}
    \BG({\BK},\omega)&=&
    [\omega - \BS(\omega) - \bm{h}^{\text{eff}}_{0}({\BK})]^{-1}.
    \label{spec_fun}
  \end{eqnarray}
  {}From this definition one can see that the two necessary
  ingredients to calculate $A({\BK},\omega)$ are (i) the Hamiltonian
  matrix $\bm{h}^{\text{eff}}_{0}({\BK})$, and (ii) the self-energy
  matrix $\BS(\omega)$ at real frequencies. Similar schemes were
  recently used by Liebsch and Lichtenstein to compute quasiparticle
  properties of Sr$_2$RuO$_4$~\cite{LDADMFTTMO1} and by Biermann~{\em
    et al.}  to describe the presence of a lower Hubbard band in
  $\gamma$-Mn \cite{Biermann01}.   Angle-resolved photoemission spectra
  of the 2D Hubbard model were also investigated by Maier~{\em et
    al.}~\cite{Maier02} in the framework of the dynamical cluster
  approximation (DCA)~\cite{DCA} {{and by Sadovskii~{\em et
        al.}~\cite{Sadovskii05} within the so-called
      DMFT+$\Sigma_{\BK}$ approach. Within DMFT the self-energy on the
      real axis was also calculated in \cite{Wannier,BluemerPhD}.}}
  
\begin{figure}
    \begin{center}
      \includegraphics[clip=true,width=0.5\textwidth,angle=270]{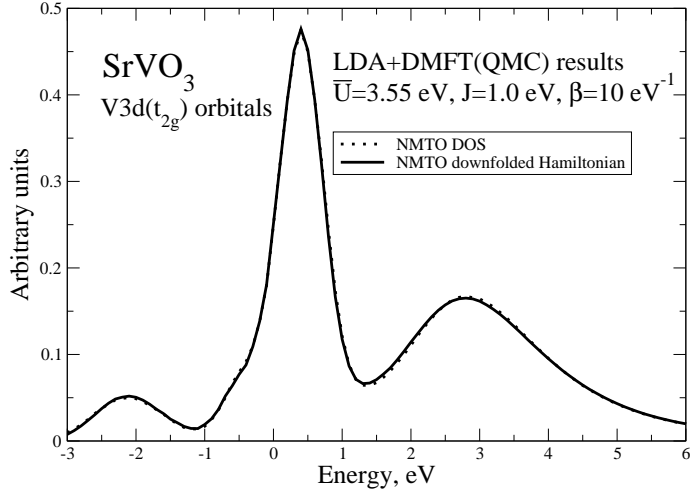}
    \end{center}
    \caption{Comparison of the LDA+DMFT(QMC) ${\BK}$-integrated 
      SrVO$_3$ spectrum of the three  $t_{2g}$ bands
      crossing the Fermi energy
      obtained by NMTO  (doted line) and 
      by the NMTO downfolded $t_{2g}$ Hamiltonian (full line), respectively. 
      The local Coulomb interaction was calculated by constrained LDA 
      as  $\bar U$=3.55 eV and $J$=1.0~eV; the
      temperature is 0.1~eV.}
    \label{qmc_nmto_dos}
  \end{figure}

  {{In our case of cubic symmetry, $\Sigma(\omega)$ is the same for
      all $t_{2g}$ orbitals.}}  Eqs.~(\ref{Ham_intg})-(\ref{spec_fun})
  are formulated in terms of a self-energy $\Sigma(\omega)$ for
  {\it{real}} frequencies $\omega$.  Since LDA+DMFT(QMC) determines
  the self-energy $\Sigma(i\omega_n)$ for Matsubara frequencies
  $i\omega_n$, the calculation of $\Sigma(\omega)$ requires a separate
  calculation.  To this end we first employ the maximum entropy method
  \cite{MEM} to obtain the ${\BK}$-integrated, spectral function
  $A(\omega)=-\frac{1}{\pi} {\rm{Im}} G(\omega)$ with
  $G(\omega)\equiv(\BG(\omega))_{mm}$, shown in Fig.
  \ref{qmc_nmto_dos}. The Kramers-Kronig relation
  \begin{eqnarray}    \label{fullGF}
    {\rm{Re}} G(\omega) =
    -\frac{1}{\pi}\int\limits_{-\infty}^{\infty}d\omega' \;\frac{{\rm{Im}}
      G(\omega')}{\omega - \omega' + i\eta}
  \end{eqnarray}
  then determines the real part of the Green function.  The complex
  Green function and the complex self-energy are related by the
  ${\BK}$-integrated Dyson Eq.~(\ref{Ham_intg}). We obtain the
  self-energy as the numerical solution of Eq.~(\ref{Ham_intg}).
  
\begin{figure}
    \begin{center}
      \includegraphics[clip=true,width=0.5\textwidth]{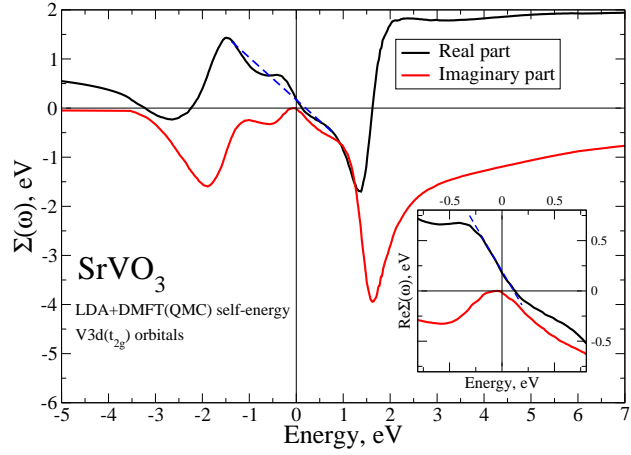}
    \end{center}
    \caption{Real (black line) and imaginary (gray line) parts of the
      LDA+DMFT(QMC) self-energy $\Sigma(\omega)$ for the vanadium
      $t_{2g}$ orbitals of SrVO$_3$ (see text). 
      The inset shows the magnification of Re$\Sigma(\omega)$ and
      Im$\Sigma(\omega)$ near the Fermi level.
      The dashed lines indicate a coarse grained linearization
      for the whole range 
      of the central peak ($-0.8\,$ to $1.4\,$eV; main figure)
      and for the strict quasiparticle regime from $-0.2\,$eV to $+0.15\,$eV (inset).}
    \label{sigma_qmc}
  \end{figure}

  {{ Fig.~\ref{sigma_qmc} presents the resulting real and imaginary
      parts of the self-energy $\Sigma(\omega)$ as a function of real
      frequencies~$\omega$.  The calculated self-energy satisfies the
      Kramers-Kronig relation}}
  \begin{equation}
    {\rm{Re}} \Sigma(\omega) = -\frac{1}{\pi} \int_{-\infty}^{\infty} \; d \omega' \frac
    {{\rm{Im}} \Sigma(\omega)}{\omega-\omega'}
    +{\rm constant}. \label{KKself}
  \end{equation}
  %% with the arbitrary constant equal to 1.24 eV.
  
  {{The self-energy is highly asymmetric with respect to the Fermi
      level, as expected for the present case of an asymmetric LDA DOS
      and 1/6 band filling.  At the energies $\omega\sim\pm$1.5 eV the
      real part of the self-energy, Re$\Sigma$, has extrema, originating
      from the crossover from the central quasiparticle peak to the lower and 
      upper Hubbard bands.  The two extrema of
      Im$\Sigma$, which coincide with zeros of Re$\Sigma$, are responsible for the strong incoherence of
      the lower and upper Hubbard bands
      (see Fig. \ref{qmc_nmto_dos}).}}
  
  {{Let us now turn from the Hubbard bands to the energy 
      regime of the central (quasiparticle) peak,
      ranging from about -0.8$\,$eV to $1.4\,$eV
      in Fig.\ \ref{qmc_nmto_dos}.
      From a coarse grained perspective, the imaginary
      part of the self-energy ${\rm Im} \Sigma(\omega)$  
      is still (relatively) small in this regime and the
      real part of the self-energy 
      can very roughly be described by
      a straight line (dashed line in Fig.\ \ref{sigma_qmc}, main panel).
      This line corresponds to a quasiparticle weight
      $Z=m^\star/m=1-\frac{\partial
        {\rm{Re}} \Sigma(\omega)}{\partial \omega}|_{\omega=0}=1.9$,
      a value which is in accord with the one   determined from
      the lowest Matsubara
      frequency $\omega_0$, i.e.,   $m^\star/m=1-\frac{{\rm{Im}}
        \Sigma(\omega_0)}{\omega_0}\approx 2$, and the estimate from
      the overall weight
      of the central (quasiparticle) peak
      (from -0.8$\,$eV to $1.4\,$eV:  $1/Z=m^\star/m\approx 2.2$).
      It is also very close to the value $m^*/m=2.2$ obtained
      in Ref.\ \cite{Pavarini03,Pavarini05} and the value $m^*/m=1.8\pm0.2$ 
      from  more recent ARPES experiments \cite{Yoshida05}.
      
      But the inset of Fig.\  \ref{sigma_qmc} reveals that, strictly speaking,
      the Fermi liquid regime with  ${\rm Im}\Sigma
      (\omega) \sim -\omega^2$ and ${\rm Re} \Sigma(\omega) \sim
      -\omega$ only extends  from -0.2$\,$ up to 0.15$\,$eV.
      The slope  of ${\rm{Re}} \Sigma(\omega)$ is steeper here than
      the slope of the coarse grained line of the main panel of  Fig.\  \ref{sigma_qmc}.
      Hence, the strict (low-energy) 
      Fermi-liquid mass enhancement is somewhat larger than $m^*/m=1.9$:
      The effective mass in this low energy (low$E$) regime is obtained as  $m^*_{{\rm low}E}/m=3$ 
       from the dashed line of the inset.}}

\begin{figure}
    \begin{center}
      \includegraphics[clip=true,width=0.5\textwidth]{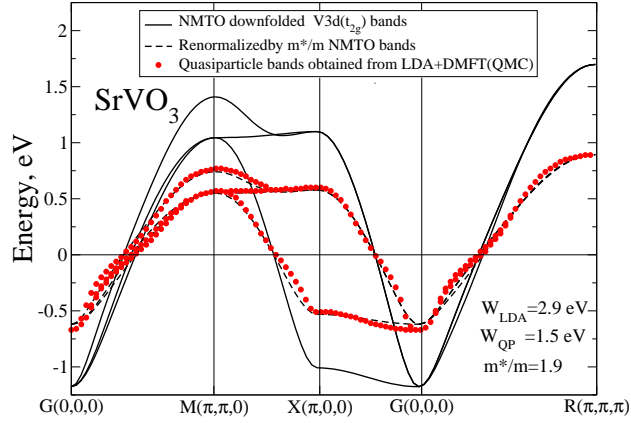}
    \end{center}
    \caption{LDA+DMFT(QMC) dispersion for  SrVO$_3$ (dots) compared 
      with LMTO  (full line) and 
      quasiparticle renormalization of the LMTO dispersion
      by $Z=1/1.9$ (dashed line).
      The ratio of bandwidths yields $1/Z=m^\star/m$=1.9.
      The dashed line represents a simple quasiparticle renormalization of
      the NMTO bands by  $1/Z=1.9$. At $\omega=-0.25\,$eV, we see a ``kink''
      in the  LDA+DMFT(QMC) dispersion
      (dots), clearly discernible as a deviation from the simple renormalized 
       LDA bands (dashed line).
}
    \label{nmto_qmc_bands}
  \end{figure}

      Next to this Fermi liquid regime, there 
      are  pronounced shoulders in ${\rm Re} \Sigma(\omega)$ at $\omega =
      -0.25$~eV and  $+0.25$~eV, with
      corresponding structures in ${\rm Im} \Sigma(\omega)$, according
      to the Kramers-Kronig relation (\ref{KKself}).    These shoulders of
      ${\rm Re} \Sigma(\omega)$ will
      become important  in the context of the quasiparticle dispersion in
      Fig.\ \ref{nmto_qmc_bands}. For  ${\rm Im} \Sigma(\omega)$,
       similar structures were reported in   \cite{BluemerPhD}, based
      on LDA+DMFT(QMC) calculations for 
      LaTiO$_3$ \cite{LDADMFTTMO2}.
       Because of the above-mentioned  shoulders in ${\rm Re} \Sigma(\omega)$, ${\rm Re} \Sigma(\omega)$ can be roughly approximated by a straight line (dashed line of the main panel Fig.\  \ref{sigma_qmc}) in the overall energy regime of the
       central quasiparticle peak.

\begin{figure}
    \begin{center}
      \includegraphics[clip=true,width=0.5\textwidth]{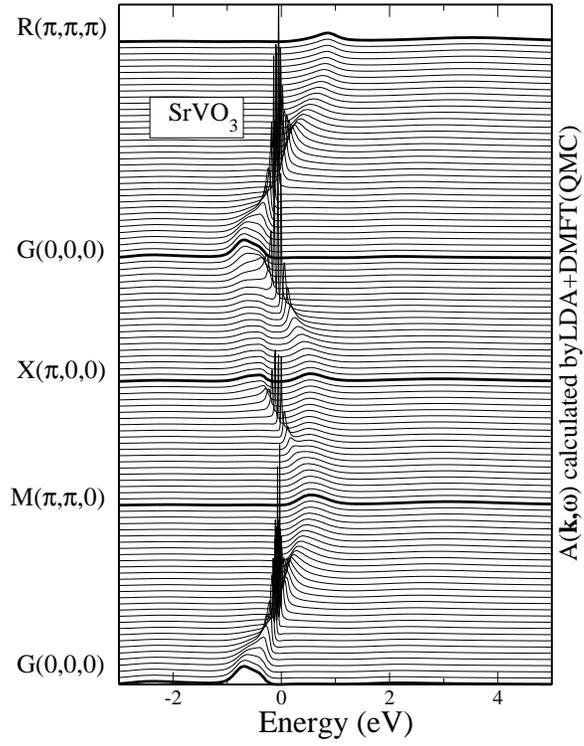}
    \end{center}
    \caption{Spectral function $A({\BK},\omega)$
      for the three V-3$d$($t_{2g}$) bands of SrVO$_3$ as calculated by LDA+DMFT(QMC).}
    \label{arpes_full}
  \end{figure}

\begin{figure}
    \begin{center}
      \includegraphics[clip=true,width=0.5\textwidth]{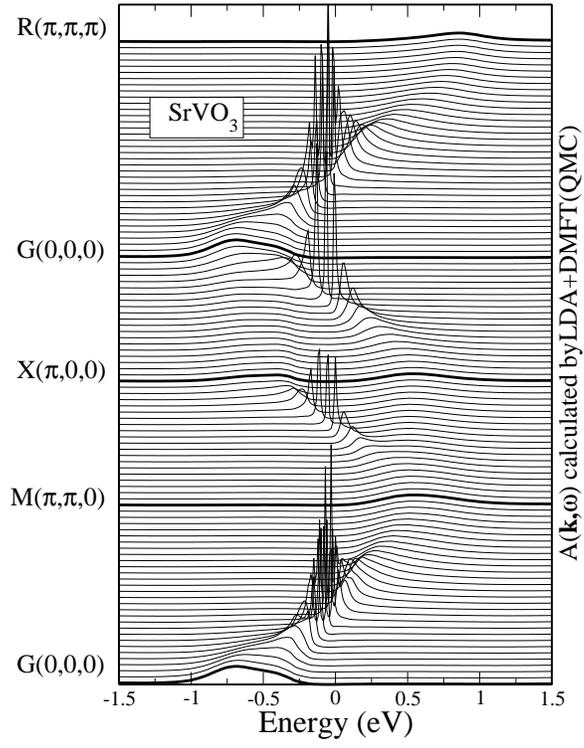}
    \end{center}
    \caption{Magnification of Fig.\ \ref{arpes_full} around
      the Fermi energy (0~eV).}
    \label{arpes_fermi}
  \end{figure}

  {{With the knowledge of the self-energy on the real axis, we are
      now in the position to calculate the {\BK}-resolved spectral
      functions Eqs.~(\ref{aofomega})-(\ref{spec_fun}) and the
      quasiparticle dispersion.}}  In Fig.\ \ref{arpes_full}, the
  LDA+DMFT(QMC) spectral functions $A({\BK},\omega)$ for SrVO$_3$ are
  presented.  In the energy regions [-3~eV, -1~eV] and [1.5~eV, 5~eV]
  there is some broad, non-dispersive spectral weight corresponding to
  the incoherent lower and upper Hubbard bands.  Around the Fermi
  energy, $A({\BK},\omega)$ shows a dispersive peak which is somewhat
  smeared out away from the Fermi energy because of lifetime effects,
  $\tau^{-1}\sim \omega^2$; the inset of Fig.\ \ref{arpes_fermi} shows
  a magnification in the vicinity of the Fermi energy.
  
  {{The {\BK}-resolved spectral functions in turn allow us to determine the
      LDA+DMFT(QMC) quasiparticle bands, which are shown as dots in
      Fig.~\ref{nmto_qmc_bands} and compared to the bare LDA bands
      (solid lines).  These dots are the maxima of the spectral
      function from Figs.~\ref{arpes_full} and~\ref{arpes_fermi}
      around the Fermi level where the quasiparticles are well
      defined.}}  They resemble the LDA dispersion, albeit
  renormalized. This is to be expected for a Fermi liquid, where
  \begin{eqnarray}
    \BG(\omega)=Z\int\limits_{\text{BZ}}\!d{\BK}\;
    [\omega  + Z \mu  - Z \bm{h}^{\text{eff}}_{0}({\BK})]^{-1}
    \label{Ham_intg2}
  \end{eqnarray}
  in the quasiparticle region.  Employing this Fermi liquid behavior,
  and using $1/Z=1.9$, we can reconstruct the band-structure directly
  from the LDA spectrum.  As seen from Fig.~\ref{nmto_qmc_bands}, the
  result (dashed curves) agrees well {{with the quasiparticle bands
      (dots).  However it should be noted that changes in slope of the
      LDA+DMFT(QMC) dispersion occur at $\omega=-0.25$~eV and (hardly discernible)
      $\omega=+0.25$~eV (see Fig.~\ref{nmto_qmc_bands}). These
      ``kinks'' \cite{Fink,Keller_abstract} stem from the shoulders in
      the real part of the self-energy (Fig.~\ref{sigma_qmc}) and will
      be discussed in detail elsewhere \cite{unpubl}. Strong interest
      in kinks of the dispersion has followed their observation in
      various high-$T_c$ superconductors \cite{hightckinks}, where
      they have been attributed mainly to phonons.  In electronic
      systems kinks in the dispersion have also been found in
      theoretical studies of the 2D Hubbard model within the
      fluctuation exchange approximation \cite{Manske} and most
      recently within the self-consistent projection operator method
      \cite{Kakehashi05}.}}
   
  When comparing with experiments, we note that for  ${\BK}$-{\it{resolved}}
  spectra the influence of PES matrix elements may be stronger than
  for the ${\BK}$-integrated spectra. Nevertheless, their inclusion
  affects the relative intensities but not their position.  {{We
      find qualitative agreement with recent ARPES dispersions
      \cite{Yoshida05}, where the renormalized band structure was
      observed directly. In particular, we see from Fig.\ 
      \ref{nmto_qmc_bands} that the bottom of the
      quasiparticle band is located at approximately $\omega=-0.7$~eV,
      in contrast to the LDA value of $\omega=-1.2$~eV.}}

  \section{Conclusion}
  \label{conclusion}
  
  In this paper, we presented LDA+DMFT(QMC) computations of
  ${\BK}$-resolved spectral functions of SrVO$_3$. The necessary
  input is an LDA-calculated Hamiltonian $\hat H_0^{\text{eff}}$ and
  the LDA+DMFT self-energy at real frequencies $\Sigma(\omega)$. We
  used the NMTO downfolding to calculate $\hat H_0^{\text{eff}}$ for
  the strongly correlated V-3$d$($t_{2g}$) orbitals of SrVO$_3$ crossing the
  Fermi energy.  This calculation gives essentially the same
  ${\BK}$-integrated spectrum as our previous calculations
  \cite{Nekrasov03,Nekrasov05x} based on the $t_{2g}$ projected DOS.

  The LDA+DMFT ${\BK}$-resolved spectral function shows two incoherent
  Hubbard bands and dispersive quasiparticle bands. The latter
  resembles the LDA dispersion which  from a coarse grained
  perspective is just normalized by  $m^\star/m$=1.9 all the way from
  the lower band edge to the  Fermi energy to the upper band
  edge of the central  (quasiparticle) peak. This   $m^\star/m$ agrees with
  ARPES experiments  \cite{Yoshida05}.
      On a finer scale we note however deviations: First,  the
      Fermi liquid
      regime only extends from -0.2$\,$eV to $0.15\,$eV, strictly speaking.
      In this low energy regime, the effective mass
      is somewhat higher  ($m^\star_{{\rm low}E}/m\approx 3$).
     Second, following this strict Fermi liquid regime the imaginary part of the self-energy stays (relatively) small, while the
     real part develops a shoulder. This shoulder
     translates into a ``kink'' in the dispersion.

  \section{Acknowledgments}
  {{We thank J.\ W.\ Allen, J.\ Fink, P.\ Fulde, D.\ Manske, K.\ Matho, and
      Y.-f.\ Yang for helpful discussions.}} This work was supported
  by Russian Basic Research Foundation grant RFFI-GFEN-03-02-39024\_a
  (VA,IN,DK), RFFI-04-02-16096 (VA,IN,DK), RFFI-05-02-17244 (IN),
  RFFI-05-02-16301 (IN) and the Deutsche Forschungsgemeinschaft
  through Sonderforschungsbereich 484 (DV,GK,MK,IN) and in part by the
  joint UrO-SO project N22 (VA,IN), the Emmy-Noether program (KH), and
  programs of the Presidium of the Russian Academy of Sciences (RAS)
  ``Quantum macrophysics'' and of the Division of Physical Sciences of
  the RAS ``Strongly correlated electrons in semiconductors, metals,
  superconductors and magnetic materials''.\ One of us (IN)
  acknowledges Dynasty Foundation and International Center for
  Fundamental Physics in Moscow program for young scientists 2005,
  Russian Science Support Foundation program for young PhD of Russian
  Academy of Science 2005.

  \pagebreak

  \pagebreak

  \pagebreak

  \pagebreak

  \pagebreak

  \pagebreak

  \pagebreak

\begin{thebibliography}{99}
    %%\vspace{-.5cm}

  \bibitem{JonesGunn}  R.~O.\ Jones and O.\ Gunnarsson, Rev.\ Mod.\ Phys.\ {\bf 61},
    689 (1989).

  \bibitem{AnLi}  V.~I.\ Anisimov, A.~I.\ Poteryaev, M.~A.\ Korotin, A.~O.\ Anokhin,
    and G.\ Kotliar, J.\ Phys.\ Cond.\ Matter {\bf 9}, 7359 (1997);
    A.~I.\ Lichtenstein, and M.~I.\ Katsnelson, Phys.\ Rev.\ B
    {\bf 57}, 6884 (1998).

  \bibitem{Held01} K.\ Held, I.~A.\ Nekrasov, N.\ Bl\"umer, V.~I.\ Anisimov,
    and D.\ Vollhardt, Int.\ J.\ Mod.\ Phys.\ B {\bf 15}, 2611 (2001).


  \bibitem{heldpsik} 
    K.\ Held, I.~A.\ Nekrasov, G.\ Keller, V.\ Eyert, N.\ Bl\"umer, A.~K.\ McMahan,
    R.~T.\ Scalettar, Th.\ Pruschke, V.~I.\ Anisimov, and D.\ Vollhardt, Psi-k Newsletter
    {\bf 56}, 65 (2003), \url{http://psi-k.dl.ac.uk/newsletters/News_56/Highlight_56.pdf}.

  \bibitem{likako} A.~I.\ Lichtenstein, M.~I.\ Katsnelson, G.\ Kotliar,
    in {\em Electron Correlations and Materials Properties 2nd ed.},
    edited by A.\ Gonis, Nicholis Kioussis, and Mikael Ciftan,
    Kluwer Academic/Plenum, p.\ 428, New York (2002), avilable as cond-mat/0112079.

  \bibitem{PT} G.\ Kotliar and D.\ Vollhardt, Physics Today \textbf{57},
    No.\ 3 (March), 53 (2004).

  \bibitem{MetzVoll89}  W.\ Metzner and D.\ Vollhardt, Phys.\ Rev.\ Lett.
    {\bf 62}, 324 (1989).

  \bibitem{vollha93}  D.~Vollhardt, in {\em Correlated Electron Systems},
    edited by V.~J.\ Emery, World Scientific, Singapore, 1993, p.~57.

  \bibitem{pruschke}  Th.\ Pruschke, M.\ Jarrell, and J.~K.\ Freericks, Adv.
    in Phys.\ {\bf 44}, 187 (1995).

  \bibitem{georges96}  A.\ Georges, G.\ Kotliar, W.\ Krauth, and M.~J.\ Rozenberg,
    Rev.\ Mod.\ Phys.\ {\bf 68}, 13 (1996).

  \bibitem{LDADMFTTMO2} I.~A.\ Nekrasov, K.\ Held, N.\ Bl\"umer,
    A.~I.\ Poteryaev,   V.~I.\ Anisimov, and D.\ Vollhardt,
    Eur.\ Phys.\ J.\ B {\bf 18}, 55 (2000).
    
  \bibitem{Held01a} K.~Held, G.~Keller, V.~Eyert, D.\ Vollhardt, and V.~I.\ Anisimov,
     Phys.\ Rev.\ Lett.\ {\bf 86}, 5345 (2001); S.\ -K.
    Mo, H.-D.\ Kim, J.~W.\ Allen, G.-H.\ Gweon, J.~D.\ Denlinger, J.-H.
    Park, A.\ Sekiyama, A.\ Yamasaki, S.\ Suga, P.\ Metcalf, and K.\ Held,
    Phys.\ Rev.\ Lett., \textbf{93}, 076404 (2004); G.\ Keller, K.\ Held,
    V.\ Eyert, D.\ Vollhardt, and V.~I.\ Anisimov, Phys.\ Rev.\ B
    \textbf{70}, 205116 (2004).

  \bibitem{Laad02b} M.~S.\ Laad, L.\ Craco, and E.\ M\"uller-Hartmann,
    Phys.\ Rev.\ Lett.\ \textbf{91}, 156402 (2003).
    
  \bibitem{Wannier} V.\ I.\ Anisimov, D.~E.\ Kondakov, A.~V.\ 
    Kozhevnikov, I.~A.\ Nekrasov, Z.~V.\ Pchelkina, J.~W.\ Allen, S.-K.\ Mo, H.-D.\ 
    Kim, P.\ Metcalf, S.\ Suga, and A.\ Sekiyama, G.\ Keller, I.\ 
    Leonov, X.\ Ren, D.\ Vollhardt, Phys.\ Rev.\ B \textbf{71}, 125119
    (2005).
    
  \bibitem{LiebschSrVO3} A.\ Liebsch, Phys.\ Rev.\ Lett.\ \textbf{90},
    096401 (2003).
    
  \bibitem{Nekrasov03} I.~A.\ Nekrasov, G.\ Keller, D.~E.\ Kondakov,
    A.~V.\ Kozhevnikov, Th.\ Pruschke, K.\ Held, D.\ Vollhardt, and
    V.~I.\ Anisimov, cond-mat/0211508, superceded by
    \cite{Sekiyama03}.

  \bibitem{Pavarini03} E.\ Pavarini, S.\ Biermann, A.\ Poteryaev, A.~I.\ Lichtenstein,
    A.\ Georges, O.~K.\ Andersen, Phys.\ Rev.\ Lett.\ \textbf{92}, 176403 (2004).

  \bibitem{Sekiyama03} 
    A.\ Sekiyama, H.\ Fujiwara, S.\ Imada, S.\ Suga, H.\ Eisaki,
    S.~I.\ Uchida, K.\ Takegahara, H.\ Harima, Y.\ Saitoh, I.~A.\ Nekrasov, G.\ Keller,
    D.~E.\ Kondakov, A.~V.\ Kozhevnikov, Th.\ Pruschke, K.\ Held, D.\ Vollhardt, 
    and V.~I.\ Anisimov, Phys.\ Rev.\ Lett.\ \textbf{93}, 156402 (2004).
    
  \bibitem{Nekrasov05x} I.~A.\ Nekrasov, G.\ Keller, D.~E.\ Kondakov,
    A.~V.\ Kozhevnikov, Th.\ Pruschke, K.\ Held, D.\ Vollhardt, and
    V.~I.\ Anisimov, cond-mat/0501240, Phys.\ Rev.\ B, in print.

  \bibitem{Pavarini05} E.\ Pavarini, A.\ Yamasaki, J.\ Nuss, and
    O.~K.\ Andersen, cond-mat/0504034.

  \bibitem{Nekrasov02}  I.~A.\ Nekrasov, Z.~V.\ Pchelkina, G.\ Keller, Th.\ Pruschke,
    K.\ Held, A.\ Krimmel, D.\ Vollhardt, and V.~I.\ Anisimov, 
    Phys.\ Rev.\ B \textbf{67}, 085111 (2003).
    
  \bibitem{LDADMFTTMO1} A.\ Liebsch and A.\ Lichtenstein, Phys.\ Rev.\ 
    Lett.\ \textbf{84}, 1591 (2000).

  \bibitem{Anisimov01} V.~I.\ Anisimov, I.~A.\ Nekrasov, D.~E.\ Kondakov,
    T.~M.\ Rice, and M.\ Sigrist, Eur.\ Phys.\ J.\ B  {\bf 25}, 191 (2002).

  \bibitem{Laad02a} L.\ Craco, M.~S.\ Laad, and E.\ M\"uller-Hartmann,
    Phys.\ Rev.\ Lett.\ \textbf{90}, 237203 (2003).

  \bibitem{Lichtenstein01}A.~I.~Lichtenstein, M.~I.~Katsnelson, and G.~Kotliar,
    Phys.\ Rev.\ Lett.\ \textbf{87}, 67205 (2001).

  \bibitem{SAVRASOV}
    S.~Y.~Savrasov, G.\ Kotliar, and E.\ Abrahams, Nature {\bf 410}, 793 (2001).

  \bibitem{SAVRASOV2}  S.~Y.\ Savrasov and G.~Kotliar, 
    Phys.\ Rev.\ B {\bf 69}, 245101 (2004).

  \bibitem{Zoelfl01}
    M.~B.\ Z\"{o}lfl, I.~A.\ Nekrasov, Th.\ Pruschke, V.~I.\ Anisimov, and
    J.~Keller, Phys.\ Rev.\ Lett.\ {\bf 87}, 276403 (2001).
    
  \bibitem{McMahan01} K.~Held, A.~K.~McMahan, and R.~T.~Scalettar,
    Phys.\ Rev.\ Lett.\ {\bf 87}, 276404 (2001).

  \bibitem{McMahan03} A.~K.~McMahan, K.~Held, and R.~T.~Scalettar,
    Phys.\ Rev.\ B {\bf 67}, 75108 (2003).
    
  \bibitem{Haule05} K.\ Haule, V.\ Oudovenko, S.~Y.\ Savrasov, and G.
    Kotliar Phys.\ Rev.\ Lett.\ \textbf{94}, 036401 (2005),
    
  \bibitem{McMahan05} A.~K.\ McMahan, cond-mat/0504380.
    
  \bibitem{Georges05} B.\ Amadon, S.\ Biermann, A.\ Georges, and F.\ Aryasetiawan, cond-mat/0504732.

  \bibitem{McMahan} A.~K.\ McMahan, R.~M.\ Martin, and S.\ Satpathy,
    Phys.\ Rev.\ B \textbf{38}, 6650 (1988).
    
  \bibitem{Hybertsen} M.~S.\ Hybertsen, M.\ Schl\"{u}ter, and N.~E.
    Christensen, Phys.\ Rev.\ B \textbf{39}, 9028 (1989).
    
  \bibitem{Knotek} M.~L.\ Knotek and P.~J.\ Feibelman, Phys.\ Rev.\ 
    Lett.\ \textbf{40}, 964 (1978).
    
  \bibitem{MTO3} O.~K.~Andersen and T.~Saha-Dasgupta, Phys.\ Rev.\ B
    {\bf 62}, R16219 (2000); O.~K.~Andersen, T.~Saha-Dasgupta,
    S.~Ezhov, L.~Tsetseris, O.~Jepsen, R.~W.\ Tank, C.\ Arcangeli, and
    G.~Krier, Psi-k Newsletter {\bf 45}, 86 (2001); O.~K.\ Andersen,
    T.~Saha-Dasgupta, and S.\ Ezhov, Bull.\ Mater.\ Sci.\ {\bf 26}, 19
    (2003).

  \bibitem{Fujimori92a}  A.\ Fujimori, I.\ Hase, H.\ Namatame, Y.\ Fujishima, Y.\ Tokura,
    H.\ Eisaki, S.\ Uchida, K.\ Takegahara, and F.~M.~F.\ de Groot,
    Phys.\ Rev.\ Lett.\ {\bf 69}, 1796 (1992).

  \bibitem{Maiti01}K.\ Maiti, D.~D.\ Sarma, M.~J.\ Rozenberg, I.~H.\ Inoue,
    H.\ Makino, O.\ Goto, M.\ Pedio, and R.\ Cimino,
    Europhys.\  Lett.\ {\bf 55}, 246 (2001).

  \bibitem{Sekiyama02} A.\ Sekiyama, H.\ Fujiwara, S.\ Imada, S.\ Suga, H.\ Eisaki,
    S.~I.\ Uchida, K.\ Takegahara, H.\ Harima, and Y.\ Saitoh, cond-mat/0206471,
    superceded by \cite{Sekiyama03}.


  \bibitem{Inoue94} I.~H.\ Inoue, I.\ Hase, Y.\ Aiura, A.\ Fujimori,
    K.\ Morikawa, T.\ Mizokawa, Y.\ Haruyama, T.\ Maruyama, and Y.\ Nishihara,
    Physica C {\bf 235-240}, 1007 (1994).

  \bibitem{Morikawa95} K.\ Morikawa, T.\ Mizokawa, K.\ Kobayashi, A.\ Fujimori, 
    H.\ Eisaki, S.\ Uchida, F.\ Iga, and Y.\ Nishihara,
    Phys.\ Rev.\ B {\bf 52}, 13711 (1995).

  \bibitem{old_experiments} Y.\ Aiura, F.\ Iga, Y.\ Nishihara, H.\ Ohnuki, H.\ Kato,
    Phys.\ Rev.\ B {\bf 47}, 6732 (1993);
    I.~H.\ Inoue, I.\ Hase, Y.\ Aiura, A.\ Fujimori, Y.\ Haruyama,
    T.\ Maruyama, and Y.\ Nishihara, Phys.\ Rev.\ Lett.\ {\bf 74}, 2539 (1995);
    I.~H.\ Inoue, O.\ Goto, H.\ Makino, N.~E.\ Hussey, and M.\ Ishikawa,
    Phys.\ Rev.\ B {\bf 58}, 4372 (1998).
    
  \bibitem{Rozenberg96} M.~J.\ Rozenberg, I.~H.\ Inoue, H.\ Makino,
    F.\ Iga, and Y.\ Nishihara, Phys.\ Rev.\ Lett.\ \textbf{76}, 4781
    (1996).

  \bibitem{Makino98} H.\ Makino, I.~H.\ Inoue, M.~J.\ Rozenberg,
    I.\ Hase, Y.\ Aiura, and S.\ Onari,  Phys.\ Rev.\ B \textbf{58}, 4384
    (1998).

  \bibitem{Eguchi05} R.\ Eguchi, T.\ Kiss, S.\ Tsuda, T.\ Shimojima, T.\ 
    Yokoya, A.\ Chainani, S.\ Shin, I.\ H.\ Inoue, T.\ Togashi, S.\ 
    Watanabe, C.\ Q.\ Zhang, C.\ T.\ Chen, M.\ Arita, K.\ Shimada, H.\ 
    Namatame, and M.\ Taniguchi, cond-mat/0504576.

  \bibitem{Yoshida05} T.\ Yoshida, K.\ Tanaka, H.\ Yagi, A.\ Ino, H.\ Eisaki, A.\ Fujimori,
    and Z.-X.\ Shen, cond-mat/0504075.

  \bibitem{Rey90} M.~J.\ Rey, P.~H.\ Dehaudt, J.~C.\ Joubert, B.\ Lambert-Andron,
    M.\ Cyrot, and F.\ Cyrot-Lackmann, J.\ Solid State Chem.\ {\bf 86}, 101
    (1990).

  \bibitem{LMTO} 
    O.\ K.\ Andersen and O.\ Jepsen, Phys.\ Rev.\ Lett.\ {\bf 53}, 2571 (1984);
    The Stuttgart TB-LMTO-ASA Program, \url{http://www.fkf.mpg.de/andersen/LMTODOC/LMTODOC.html}.

  \bibitem{comment} Our results are in agreement with LDA calculations within the
     LMTO-ASA method by S. Itoh, Solid State Communications, \textbf{88}, 525 (1993), and
     those  by  J.\ Takegahara, 
     Electron Spectrosc.\ Relat.\ Phenom.\ \textbf{66}, 303 (1995)
     in the basis of augmented plane waves (APW).\ For our
    purposes, i.e., for the combination of the LDA band structure with
    DMFT~\cite{AnLi}, the atomic-like wave functions basis set of the MTO
    method is however more appropriate. 

  \bibitem{Lambin84} Ph.\ Lambin and J.~P.\ Vigneron, Phys.\ Rev.\ B {\bf 29}, 3430 (1984);
    O.\ Jepsen and O.\ K.\ Andersen, Solid State Commun.\ {\bf 9}, 1763 (1971);
    P.\ E.\ Bl\"ochl, O.\ Jepsen, and O.\ K.\ Andersen, Phys.\ Rev.\ B {\bf 49}, 16223 (1994).

  \bibitem{Zoelfl00} M.~B.\ Z\"{o}lfl, Th.\ Pruschke, J.\ Keller,
    A.~I.\ Poteryaev, I.~A.\ Nekrasov, and V.~I.\ Anisimov,
    Phys.\ Rev.\ B \textbf{61}, 12810 (2000).

  \bibitem{Gunnarsson89} O.\ Gunnarsson, O.~K.\ Andersen, O.\ Jepsen, and J.\ Zaanen,
    Phys.\ Rev.\ B \textbf{39}, 1708 (1989).

  \bibitem{QMC} J.~E.\ Hirsch and R.~M.\ Fye, Phys.\ Rev.\ Lett.\ \textbf{56}, 2521
    (1986).\ For multi-band QMC within DMFT see~\cite{Held01}.

  \bibitem{MEM} For a review of the maximum entropy method see
    M.~Jarrell and J.~E.\ Gubernatis, Physics Reports {\bf 269}, 133 (1996).

  \bibitem{Biermann01}S.\ Biermann, A.\ Dallmeyer, C.\ Carbone, W.\ Eberhardt,
    C.\ Pampuch, O.\ Rader, M.~I.\ Katsnelson, and A.~I.\ Lichtenstein,
    JETP Letters {\bf 80}, 612 (2005).

  \bibitem{Maier02} Th.~A.\ Maier, Th.\ Pruschke, and M.\ Jarrell,
    Phys.\ Rev.\ B {\bf 66}, 075102 (2002).

  \bibitem{DCA} For a review of DCA see Th.\ Maier, M.\ Jarrell, Th.\ Pruschke,
    M.~H.\ Hettler, ``Quantum Cluster Theories'', cond-mat/0404055,
    (submitted to Rev.\ Mod.\ Phys.).
    %%Rev.\ Mod.\ Phys., in print.

  \bibitem{Sadovskii05} M.~V.\ Sadovskii, I.~A.\ Nekrasov, E.~Z.\ Kuchinskii,
    Th.\ Pruschke, V.~I.\ Anisimov, cond-mat/0502612; E.~Z.\ Kuchinskii, I.~A.\ Nekrasov,
    and M.~V.\ Sadovskii, cond-mat/0506215, 
    JETP Letters, in print.
    
  \bibitem{BluemerPhD} N.\ Bl\"{u}mer, Ph.~D.\ thesis, Universit\"{a}t
    Augsburg, 2002.

  \bibitem{Fink} {{We thank  J.\ Fink for useful
        discussions regarding these kinks during the 8th
        Japanese-German Symposium on \emph{Competing Phases in Novel
          Condensed Matter Systems}, Lauterbad, August 2004, where our
        results were first presented.  We also thank him
         and P.\ Fulde for useful discussions at the
        International Workshop on \emph{Strong Correlations and ARPES:
          Recent Progress in Theory and Experiment}, Max-Planck
        Institute for Physics of Complex Systems, Dresden, April
        2005.}}

  \bibitem{Keller_abstract} {{G.\ Keller, D.~E.\ Kondakov, I.\ Nekrasov,
    K.\ Held, T.\ Pruschke, V.~I.\ Anisimov, and D.\ Vollhardt,
    International Conference on Strongly Correlated Electron Systems
    SCES '04, Abstract Booklet WE-TMO3-55, p.131; G.\ Keller, V.\ I.\ 
    Anisimov, D.~E.\ Kondakov, A.~V.\ Kozhevnikov, I.~A.\ Nekrasov,
    Z.~V.\ Pchelkina, I.\ Leonov, X.\ Ren, and D.\ Vollhardt,
    Verhandl.\ DPG (VI) 40, 2/2005, TT 16.87, p.\ 580.}}

  \bibitem{unpubl} {{I.\ Nekrasov \emph{et al.}, to be published.}}
    
  \bibitem{hightckinks} 
    A.\ Lanzara {\em et al.},
% P.~V.\ Bogdanov, X.~J.\ Zhou,
%    S.~A.\ Kellar, D.~L.\ Feng, E.~D.\ Lu, T.\ Yoshida, H.\ Eisaki,
%    A.\ Fujimori, K.\ Kishio, J.-I.\  Shimoyama, T.\ Noda, S.\ Uchida,
%    Z.\ Hussain, and Z.-X.\ Shen, 
    Nature {\bf 412}, 510 (2001).
    For reviews see T.\ Cuk, D.~H.\ Lu, X.~J.\ Zhou, Z.-X.\ Shen,
    T.~P.\ Deveraux, and N.\ Nagaosa, phys.\ stat.\ sol.\ (b) {\bf
      242}, 11 (2005); A.\ Damascelli, Z.\ Hussain, and Z.-X.\ Shen,
    Rev. Mod. Phys. {\bf 75}, 473 (2003).
    
  \bibitem{Manske} D.\ Manske, I.\ Eremin, and K.\ H.\ Bennemann,
    Phys.\ Rev.\ Lett.\ {\bf 87}, 177005 (2001); Phys.\ Rev.\ B {\bf
      67}, 134520 (2003).

  \bibitem{Kakehashi05}
    Y.\ Kakehashi and P.\ Fulde, cond-mat/0507564.

  \end{thebibliography}
\end{document}